\documentclass[english, prl, showpacs, superscriptaddress,twocolumn]{revtex4-1}
\usepackage[T1]{fontenc}
\usepackage{amsmath}
\usepackage{amssymb}
\usepackage{graphicx}
\usepackage{esint}
\usepackage{bm}

\makeatletter
\@ifundefined{textcolor}{}
{%
 \definecolor{BLACK}{gray}{0}
 \definecolor{WHITE}{gray}{1}
 \definecolor{RED}{rgb}{1,0,0}
 \definecolor{GREEN}{rgb}{0,1,0}
 \definecolor{BLUE}{rgb}{0,0,1}
 \definecolor{CYAN}{cmyk}{1,0,0,0}
 \definecolor{MAGENTA}{cmyk}{0,1,0,0}
 \definecolor{YELLOW}{cmyk}{0,0,1,0}
}

\usepackage{color}
\usepackage{ulem}
\makeatother

\usepackage{babel}
\begin{document}

\title{How nonlinear interactions challenge the three-dimensional Anderson transition}

\author{Nicolas Cherroret}

\affiliation{Laboratoire Kastler-Brossel, UPMC-Paris 6, ENS, CNRS; 4 Place Jussieu,
F-75005 Paris, France}

\author{Beno\^it Vermersch}

\altaffiliation{Present address: Institute for Quantum Optics and Quantum Information of the Austrian Academy of Sciences, A-6020 Innsbruck, Austria}

\selectlanguage{english}

\author{Jean Claude Garreau}

\affiliation{Laboratoire de Physique des Lasers, Atomes et Mol\'ecules, Universit\'e Lille 1 Sciences et Technologies, CNRS; F-59655 Villeneuve d'Ascq Cedex, France}

\author{Dominique Delande}

\affiliation{Laboratoire Kastler-Brossel, UPMC-Paris 6, ENS, CNRS; 4 Place Jussieu,
F-75005 Paris, France}

\date{\today}
\begin{abstract}
In disordered systems, our present understanding of the Anderson transition is hampered by the possible presence of interactions between particles. We demonstrate that in boson gases, even weak interactions deeply alter the very nature of the Anderson transition. While there still exists a critical point in the system, below that point a novel phase appears,  displaying a new critical exponent, subdiffusive transport and a breakdown of the one-parameter scaling description of Anderson localization.
\end{abstract}
\maketitle

When a quantum particle propagates in a three-dimensional (3D) disordered medium, a spatial localization of its wave function occurs when the particle's energy is below a critical value: This is the Anderson transition, a phase transition separating a regime where transport is diffusive from a regime where transport is inhibited~\cite{Anderson58}. Since Anderson localization is fundamentally an interference phenomenon, it is not specific of quantum particles and might exist for any kind of wave \cite{Lagendijk09}. This observation led to first demonstrations of localization of light in dielectric materials~\cite{Schwartz07,Sperling12} and of acoustic waves in elastic networks~\cite{Hu08}. In atomic physics, experiments provided some evidence for localization of ultracold atoms in 3D disordered potentials~\cite{
Jendrzejewski12,Kondov11, Semeghini14}. In parallel, the Anderson transition and its critical properties were thoroughly investigated in experiments on cold atoms subjected to quasiperiodically modulated laser pulses, a system called ``kicked rotor'' which emulates the physics of a true, 3D disordered medium~\cite{Casati89, Chabe08,Lemarie10}.

For ultracold atoms in a disordered potential, interactions between particles complicate this picture. Research on one-dimensional (1D) disordered systems of bosons evidenced that Anderson localization can be strongly affected by interactions, leading to new phases at zero~\cite{Pasienski10,Deissler10} or finite~\cite{Aleiner10} temperature or to changes in the dynamics of wave packets~\cite{Pikovsky08,Kopidakis08}. In contrast, the role of interactions in the 3D Anderson transition is unknown. In this Letter we develop a generalized self-consistent theory (SCT) of localization for a spreading, weakly interacting, boson gas~\cite{Vollhardt92}. It reveals that interactions deeply affect the nature of the Anderson transition: In the interacting gas a transition still exists and the gas dynamics at and above the critical point are qualitatively 
unchanged. On the contrary, below the critical point arbitrarily weak interactions destroy Anderson localization, which is then replaced by subdiffusion. At the critical point the subdiffusion coefficient diverges algebraically, with a critical exponent different from that of the Anderson transition.

To unveil the physics of a 3D boson gas in the presence of both disorder and weak repulsive interactions, we analyze the spreading of a narrow wave packet in a static random potential. Interactions are treated within a mean field approach: The matter wave is described by a single-particle wave function $\Psi$ which fulfills the Gross-Pitaevskii equation 
\begin{equation}
i\hbar\dfrac{\partial\Psi}{\partial t}=\dfrac{-\hbar^{2}}{2m}\boldsymbol{\nabla}^{2}\Psi+V(\textbf{r})\Psi+g|\Psi|^{2}\Psi.\label{GPE}
\end{equation}
Short-range interactions are described by the nonlinear potential $g|\Psi|^{2}$. This makes our analysis also relevant for the context of electromagnetic waves in disordered media described by the Helmholtz equation \cite{Schwartz07, Hartung08, Lahini08, Wellens09} including a cubic (Kerr) nonlinearity. For simplicity we consider a white-noise model of disorder $V(\textbf{r})$ for which the amplitude $U\rightarrow\infty$ and correlation length $\sigma\rightarrow0$ with $U^{2}\sigma^{3}=\text{constant}$ \cite{footnote1}. For this model the disorder is characterized by a unique energy scale $E_{c}\sim(U^{2}\sigma^{3})^2m^{3}/\hbar^{6}$ which coincides (up to a prefactor of the order of unity) with the critical energy where the Anderson transition occurs~\cite{Skipetrov08}. $g|\Psi|^{2}$ ($g>0$) defines another energy scale, the interaction energy per particle, which we assume smaller than the 
kinetic $E$ and the disorder 
$E_{c}$ energies:
\begin{equation}
g|\Psi|^{2}\ll E,E_{c}.\label{weak_gn}
\end{equation}
We develop a generalized SCT of localization including the nonlinear potential $g|\Psi|^2$.
The essential idea of this approach is that weak interactions in the sense of Eq.~(\ref{weak_gn}) break the equivalence between Diffuson and Cooperon. For $g=0$ the Diffuson describes configurations where the amplitudes $\Psi$ and $\Psi^{*}$ follow an identical multiple scattering path. It 
controls the diffusive spreading of the wave packet that occurs for $E\gg E_{c}$, with a mean square width $\langle\textbf{r}^{2}\rangle\sim D_0t$ ($D_0$ is the diffusion coefficient). The Cooperon describes interference configurations where $\Psi$ and $\Psi^{*}$ follow the same path but in opposite directions. It renormalizes $D_0$ and, when treated self-consistently, allows for a description of Anderson localization yielding $\langle\textbf{r}^{2}\rangle\sim\xi^{2}$ for $E\lesssim E_{c}$ \cite{Vollhardt92} ($\xi$ is the localization length). Exactly at the critical point $E=E_{c}$ \cite{footnote2}, the system is intermediate between a conductor and an insulator, and $\langle\textbf{r}^{2}\rangle\sim t^{2/3}$ \cite{Shapiro82}. For a system with time-reversal symmetry, Diffuson and Cooperon are equal when $g=0$ but not when $g\ne0$. The reason is that unlike the Diffuson, the Cooperon is modified by multiple-wave contributions due to the potential $g|\Psi|^2$. 
It was shown that for $E\gg E_c$ the net result of these contributions is a \emph{dephasing} $\phi\propto g|\Psi|^2$
affecting the Cooperon~\cite{Hartung08, Wellens09}. We generalize this description to the wave packet scenario, and beyond the critical point by means of the self-consistent scheme of \cite{Vollhardt92}. The non-equivalence between Diffuson $P$ and Cooperon $P'$ requires to introduce a \emph{set} of self-consistent equations, with a renormalized diffusion coefficient $D$ for $P$ and another $D'$ for $P'$~\cite{Eckert}:
\begin{equation}
\label{NL_SCE}
\begin{array}{l}
\left\{ \begin{array}{l}
\left[-i\omega-D\boldsymbol{\nabla}^2\right]P(\textbf{r}',\textbf{r},\omega)=\delta(\textbf{r}-\textbf{r}')\\
\dfrac{1}{D}=\dfrac{1}{D_{0}}+\dfrac{1}{\pi\rho\hbar D_{0}}P'(\textbf{r},\textbf{r},\omega)\\
\left[-i\omega-D'\boldsymbol{\nabla}^2-i\dfrac{g}{\hbar}n(\textbf{r},\omega)*\right]R'(\textbf{r}',\textbf{r},\omega)=\dfrac{\gamma}{\tau}\delta(\textbf{r}-\textbf{r}')\\
\dfrac{1}{D'}=\dfrac{1}{D_{0}}+\dfrac{1}{\pi\rho\hbar D_{0}}P(\textbf{r},\textbf{r},\omega),
\end{array}\right.
\end{array}
\end{equation}
where $P'(\textbf{r}',\textbf{r},t)=(\tau/\gamma)\Re\left[R'(\textbf{r}',\textbf{r},t)\right]$ \cite{footnote3}. $\tau$ is the scattering time and $\gamma=\hbar/(2\pi\rho\tau)$ with $\rho$ the density of states per unit volume. Eqs. (\ref{NL_SCE}) rely on an hydrodynamic approximation and thus hold at long times $t\gg\tau$ and large distances $|\textbf{r}-\textbf{r}'|\gg\ell$ only ($\ell$ is the mean free path). Furthermore, Eqs. (\ref{NL_SCE}) describe the dynamics at a single energy $E$, any mixing of energies due to interactions being neglected \cite{Schwiete10}. This approximation is valid in the limit $g|\Psi|^2\ll E$ 
where the dephasing mechanism is dominant \cite{footnote4}. Note finally that for $g=0$ the SCT is known to neglect fluctuations around the critical point, and consequently cannot predict correct values of critical exponents \cite{Vollhardt92}. As shown below this conclusion applies to the 
interacting case as well. Eqs. (\ref{NL_SCE}) highlight the dephasing $\phi\propto gn$ by which interactions compete with localization: $\phi$ alters the Cooperon $P'$ via a frequency convolution $*$ reminiscent of the multiplicative potential $g|\Psi|^2$ in Eq. (\ref{GPE}). $D'$ and $P'$ should be seen as intermediate variables entering the calculation of $D$ and $P$, the relevant quantities that control the disorder averaged density: $n(\textbf{r},t)=\int d^{3}\textbf{r}'P(\textbf{r}',\textbf{r},t)|\Psi(\textbf{r}',t=0)|^2$. As expected they disappear in the limit $g=0$ where the equations for $(P,D)$ and $(P',D')$ become identical \cite{Vollhardt92}.
\begin{figure}
\includegraphics[width=8.7cm]{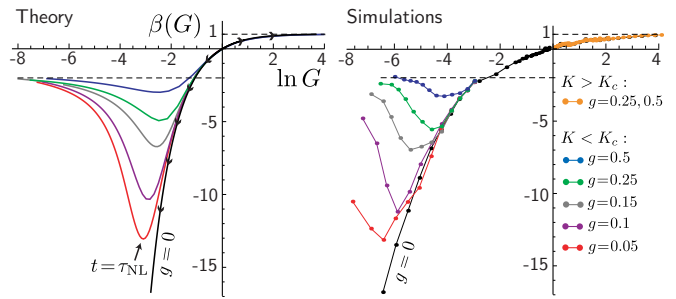}
\caption{\label{beta} (color online). Left: scaling function $\beta$ predicted by the SCT for $g=0$ (black curve) and for five finite values of $g/(D_{0}\ell\hbar)$ [colored curves, $g/(D_{0}\ell\hbar)$ increases from bottom to top]. Right: corresponding numerical simulations of the QPKNR [Parameters are $\omega_2=2\pi\sqrt{5}$, $\omega_3=2\pi\sqrt{13}$, $\epsilon=0.44$ and $\hbar=2.89$, leading to $K_{c}=6.40$. $G$ is averaged over $400-1000$ initial conditions].}
\end{figure}

We solve Eqs. (\ref{NL_SCE})  for a narrow wave packet centered at $\textbf{r}'=0$ [$n(\textbf{r},t)\simeq P(0,\textbf{r},t)$] by Fourier transforming space-dependent variables. The presence of the convolution in $\omega$ requires a special treatment that will be discussed elsewhere \cite{Tobepub}. We obtain implicit formulas for $D$ and $P$ which we solve numerically to access the scaling function
\begin{equation}
\beta(G)=\dfrac{d\ln G(L)}{d\ln L}.\label{betaL}
\end{equation}
When $g=0$, the evolution of a disordered system maps as a motion on the curve $\beta(G)$ which is controlled by a single parameter, the conductance $G$, function of the system size $L$~\cite{Abrahams79}. This description can be extended to the scenario of wave packet spreading: The role of the system size is played by the wave packet width, $L\equiv\langle\textbf{r}^{2}\rangle^{1/2}=[\int d^3\textbf{r}\, r^2P(0,\textbf{r},t)]^{1/2}$, and the ``conductance'' is defined as $G\propto L\times D=\langle\textbf{r}^{2}\rangle^{3/2}/t$, in analogy with its expression for an electronic conductor~\cite{Abrahams79}. The results for $\beta(G)$ are shown in the left panel of Fig.~\ref{beta}. For $g=0$ (black curve) $\beta$ is monotonic, positive in the diffusion regime, negative in the localization regime and zero at the critical point as expected \cite{Vollhardt92}. Far in the diffusive 
phase $\beta(G)\rightarrow1$ since $\langle\textbf{r}^{2}\rangle\sim D_{0}t$. 
Fig.~\ref{beta} also shows $\beta$ for $g\ne0$, which constitutes the main result of this Letter. First, we observe that for all curves the point where $\beta$ vanishes is still present, and unmodified. In other words, for weak interactions a phase transition still exists at $E=E_c$. Above this point $\beta$ remains unaffected by interactions. This can be understood by noting that $\phi\propto gn(\textbf{r},t)$ quickly tends to zero as the wave packet gets diluted by diffusion. In contrast, below the critical point the shape of $\beta$ is totally different when $g\ne0$: Interactions give rise to a \emph{minimum} associated with a breakdown of monotonicity, and $\beta(G\to0)\to-2$.
Coming back to the definition $G=\langle\textbf{r}^{2}\rangle^{3/2}/t$, we see that this limit corresponds to $\langle\textbf{r}^{2}\rangle\sim t^{2/5}$, i.e. to a \emph{subdiffusive phase}. This result can also be obtained directly by solving Eqs. (\ref{NL_SCE}) analytically for $\omega\to 0$. It is reminiscent of the 1D subdiffusion observed numerically~\cite{Pikovsky08,Kopidakis08, Shepelyansky93, Garcia09}. Within our SCT, subdiffusion results from a trade-off between interference and dephasing: On the one hand, interference localizes the packet and thus reinforces interactions by preventing $\phi$ from decreasing to zero with time. On the other hand, interactions delocalize the packet, which makes $\phi$ decrease and in turn reinforces interference.
The minimum of the curves corresponds to a characteristic time $t\sim\tau_{\text{NL}}\equiv \hbar \xi^3/g$, associated with a crossover between a transient localization regime and the asymptotic subdiffusion \cite{footnote6}.

To confirm these predictions, we carry out numerical simulations of wave packet spreading. Instead of solving directly the challenging 3D problem~(\ref{GPE}), we resort to the simpler model of the quasiperiodic kicked nonlinear rotor (QPKNR) \cite{Shepelyansky93, Garcia09} which 
has the Hamiltonian
\begin{eqnarray}
H&=&p^{2}/2+K\cos x(1+\epsilon\cos\omega_{2}t\cos\omega_{3}t)\sum_{n}\delta(t-n)\nonumber\\
&&+ g\hbar/(2\pi)|\Psi(p,t)|^{2}
\label{KK_H}
\end{eqnarray}
in dimensionless units. For $g=0$ and $\omega_{2}$, $\omega_{3}$, $\pi$ and $\hbar$ incommensurate, Eq. (\ref{KK_H}) can be formally mapped onto a 3D Anderson model, and displays the Anderson transition between localization for $K<K_{c}$ and diffusion for $K>K_{c}$ in \emph{momentum} space~\cite{Fishman82, Casati89,Chabe08,Lemarie10}. As compared to Eq. (\ref{GPE}), $K$ is now the control parameter instead of $E$, and the wave packet mean square width is $\langle p^{2}\rangle$ instead of $\langle\textbf{r}^{2}\rangle$, with $G=\langle p^{2}\rangle^{3/2}/t$. For $g\ne0$ the equivalence with Eq. (\ref{GPE}) strictly speaking disappears as the nonlinear term in Eq. (\ref{KK_H}) acts in a 1D space. As shown below however,  this difference does not qualitatively affect the physics, while solving numerically the QPKNR rather than Eq.~(\ref{GPE}) is much cheaper. We show in the right panel of Fig.~\ref{beta} the numerical $\beta(G)$. 
Clearly, we recover the behavior found theoretically: Interactions have no effect at and above the critical point (orange points, $g\ne0$, coincide with black points, $g=0$), while below $\beta$ has a minimum and an asymptotic value $\beta(G\rightarrow0)\simeq-2$.

\begin{figure}
\includegraphics[width=8.7cm]{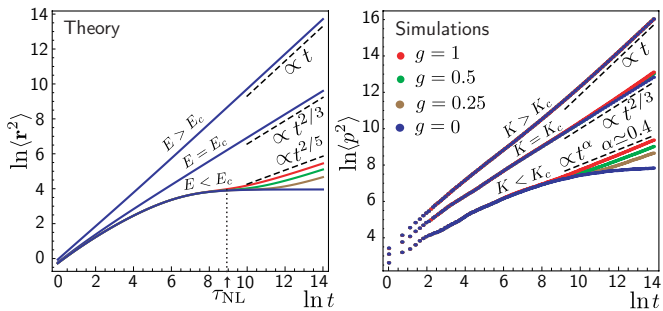}
\caption{\label{r2_t_plots} (color online). Left: time dependence of $\langle\textbf{r}^{2}\rangle$ predicted by the SCT for $E>E_{c}$, $E<E_{c}$ and $E=E_{c}$ [colored curves correspond to different values of $g/(D_{0}\ell\hbar)$, increasing from bottom to top]. Only for $E<E_{c}$ do interactions qualitatively modify the long-time asymptotics, indicated by dashed lines. Right: time dependence of $\langle p^{2}\rangle$ for $K=5$, $K=6.4=K_{c}$ and $K=7.6$, obtained from numerical simulations of the QPKNR.}
\end{figure}
We show in the left panel of Fig.~\ref{r2_t_plots} the explicit time dependence of $\langle\textbf{r}^{2}\rangle$ predicted by Eqs. (\ref{NL_SCE}). The results confirm the picture already outlined in Fig.~\ref{beta}: Interactions have no visible effect for $E\geqslant E_{c}$. In particular, the asymptotic laws $\langle\textbf{r}^{2}\rangle\sim t$ (for $E>E_{c}$) and $\langle\textbf{r}^{2}\rangle\sim t^{2/3}$ (for $E=E_{c}$) are robust against interactions. On the other hand, for $E<E_{c}$ a deviation from localization takes place at $t\sim\tau_{\text{NL}}$, corresponding to the minimum of $\beta$, see Fig.~\ref{beta}. At long times the system is subdiffusive  with $\langle\textbf{r}^{2}\rangle\propto t^{2/5}$, corresponding to the limit $\beta(G\to0)\simeq-2$ in Fig.~\ref{beta}. These predictions are confirmed by simulations of $\langle p^{2}\rangle$ using the QPKNR and shown in the right panel of Fig.~\ref{r2_t_plots}: Again interactions do not affect the dynamics 
for $K\geqslant K_{c}$ and turn localization into subdiffusion for $K<K_{c}$. Our simulations as well as our SCT do not indicate any threshold for $g$: Localization is destroyed even for arbitrarily small $g$, provided that one can reach times $t\gg \tau_\text{NL}=\hbar \xi^3/g$.

As underlined above, our disorder theory (\ref{NL_SCE}) is strictly speaking not directly applicable to the QPKNR which pertains to a 1D configuration space. While we expect this difference to be crucial for a precise determination of the subdiffusion exponent \cite{Shepelyansky93, Kopidakis08, Pikovsky08, Ermann14}, we see from Figs. \ref{beta} and \ref{r2_t_plots} that it can still be used for a semi-quantitative description of $\beta(G)$ and of subdiffusion at long times. 
\begin{figure}
\begin{centering}
\includegraphics[width=6.5cm]{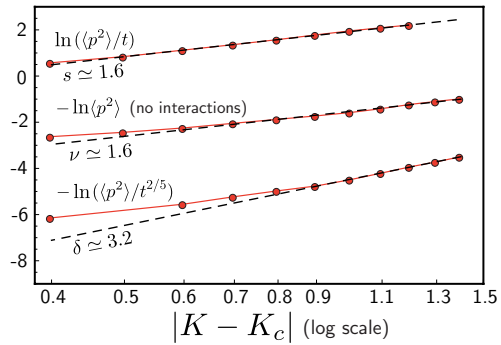}
\par\end{centering}
\caption{\label{p_kmoinsKc}
Numerical test of the scaling laws analog to Eqs.~(\ref{r2_diff}) and~(\ref{r2_nonlinear}) for the QPKNR (upper and lower curves, $K>K_{c}$ and $K<K_{c}$ respectively). Time is fixed to $10^{6}$ and $g$ to $0.5$. For comparison a plot of $\langle p^{2}\rangle$ for $K<K_{c}$ and $g=0$ is also shown, and well reproduces Eq.~(\ref{r2_loc}) (this curve is shifted upwards for clarity). Fits of the observed linear behaviors (dashed lines) give access to the critical exponents $s$, $\nu$ and $\delta$. Deviations near $K=K_c$ are finite time effects.
}
\end{figure}

Let us now analyze more carefully the physics of the interacting system around $E_{c}$, and for this purpose first recall the essential properties of the Anderson transition that occurs when $g=0$. The latter shows up at  $t\to\infty$~\cite{Shapiro82} and is characterized by two critical exponents $s$ and $\nu$ which respectively control the vanishing of the diffusion coefficient for $E\to E_{c}^{+}$, 
\begin{equation}
\dfrac{\langle\textbf{r}^{2}\rangle}{t}\underset{t\to\infty}{\sim}D\propto(E-E_{c})^{s},\label{r2_diff}
\end{equation}
and the divergence of the localization length for $E\to E_{c}^{-}$,
\begin{equation}
\sqrt{\langle\textbf{r}^{2}\rangle}\underset{t\to\infty}{\sim}\xi\propto\dfrac{1}{(E_{c}-E)^{\nu}}.\label{r2_loc}
\end{equation}
In 3D, $s$ and $\nu$ are equal (Wegner law)~\cite{Wegner76} and universal~\cite{Abrahams79} but their value is nontrivial. For a spinless, time-reversal invariant system, numerical simulations revealed that $\nu=s\simeq1.58$~\cite{Slevin01}, and the same value within error bars was measured in an experimental realization of the atomic kicked rotor \cite{Lopez12}. For $g\ne0$, a phase transition still exists at $E=E_{c}$ because: (i) the $\beta$ function changes sign (Fig.~\ref{beta}) and (ii) a small change in $E$ around $E_{c}$ leads to qualitatively different dynamics at long times (Fig.~\ref{r2_t_plots}). However, this transition is fundamentally different from the Anderson transition since localization has been replaced by subdiffusion. Solving Eqs. (\ref{NL_SCE}) for $\omega\to0$ we find that for $g\ne0$ $D$ still vanishes algebraically at $E=E_{c}^{+}$ according to Eq.~(\ref{r2_diff}), interactions only 
affecting $D$ 
at short times. For $E\to E_{c}^{-}$ there is no longer localization but subdiffusion, and Eqs. (\ref{NL_SCE}) predict that the \emph{subdiffusion coefficient} is critical: 
\begin{equation}
D_{\alpha}\equiv\dfrac{\langle\textbf{r}^{2}\rangle}{t^{\alpha}}\underset{t\to\infty}{\sim}\dfrac{g^{\alpha}}{(E_{c}-E)^{\delta}}.\label{r2_nonlinear}
\end{equation}
In this relation $\alpha=2/5$ is the subdiffusion exponent and a critical exponent $\delta$ appears. Its value is related to $s$ through a generalized Wegner law: For 3D disorder Eqs. (\ref{NL_SCE}) give $\delta=2\alpha s$, and we infer $\delta=6\alpha s$ by dimensional analysis for the QPKNR. As for $g=0$ we expect this Wegner law to be correctly captured by the SCT \cite{Vollhardt92}, even though this is not the case for the value of the critical exponents themselves \cite{footnote5} (the SCT gives $s=1$). To test the critical character of $D$ and $D_\alpha$ predicted by the SCT and to access the value of $\delta$ and $s$ for $g\ne0$, we numerically compute, using the QPKNR, $\langle p^{2}\rangle/t$ as a function of $K-K_{c}$ (for $K>K_c$) and $\langle p^{2}\rangle/t^{2/5}$ as a function of $K_c-K$ (for $K<K_c$), see Fig. \ref{p_kmoinsKc} (upper and lower curves). Algebraic scalings are well visible and linear fits give $s=1.6\pm0.1$ and $\delta=3.2\pm0.7$. These results reveal that $s$ is not affected by 
interactions and are compatible with the prediction $\delta=6\alpha s$. 

\begin{figure}
\begin{centering}
\includegraphics[width=7cm]{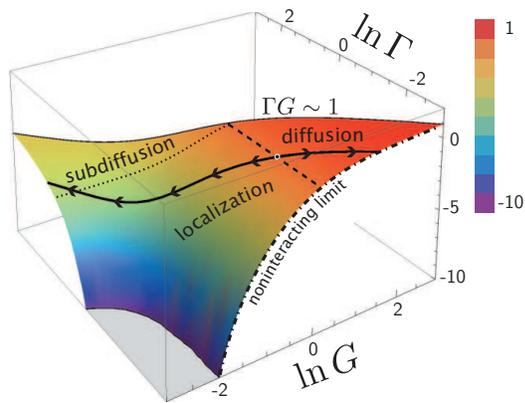}
\par\end{centering}
\caption{\label{Phase_diagram} (color online). Two-parameter scaling function $\beta(G,{\Gamma})$ derived from the SCT.
The solid curve indicates the trace of the $\beta$ functions of Fig.~\ref{beta}, corresponding to fixed values of $\lambda=g/(D_{0}\ell\hbar)$. The dashed line gives the position of the critical point $G\simeq1$ (where $\beta=0$), and the dotted curve indicates the crossover $\Gamma\simeq1$ between localization and subdiffusion. The dotted-dashed curve shows $\beta$ for $g\to0$. 
}
\end{figure}
Fig.~\ref{beta} shows that different $g$ generate different functions $\beta$, which suggests that a single-parameter description of the system is no longer possible when $g\ne0$. We find that corrections due to interactions systematically appear in the SCT as terms proportional to $\Gamma=\lambda/G$, where $\lambda=g/(D_{0}\ell\hbar)$. This naturally leads us to define $\Gamma$ as an additional scaling parameter, yielding a \emph{two-parameter scaling theory} characterized by $\beta=\beta(G,\Gamma)$ shown in Fig.~\ref{Phase_diagram}. The three regions of diffusion, Anderson localization and subdiffusion are highlighted. The solid curve indicates the trace of the $\beta$ functions of Fig.~\ref{beta}, corresponding to fixed values of $\lambda$. The dashed line gives the position of the critical point $G\simeq1$ where $\beta$ vanishes. The dotted curve 
shows the boundary $\Gamma\simeq1$ which, unlike the line $G\simeq1$, is not associated with a phase transition ($\beta\ne0$) but with the crossover between localization and subdiffusion at $t\sim\tau_{\text{NL}}$. 
Finally, for $\lambda=\Gamma G>1$ interactions start to affect the diffusion process itself, a regime where our assumption~(\ref{weak_gn}) no longer holds.

In conclusion, we demonstrated that weak bosonic interactions drive the 3D Anderson transition to a subdiffusion-diffusion transition characterized by novel critical properties and by a two-parameter scaling theory. Mathematically this transition replaces the Anderson transition even for arbitrarily small interactions, but is visible only after a correspondingly long time. In practice it is the time scale of the experiment as compared to $\tau_{\text{NL}}=\hbar\xi^{3}/g$ that decides which transition will be observed. Recent observations of 1D subdiffusion~\cite{Lucioni11} and 3D Anderson localization~\cite{Jendrzejewski12,Kondov11,Semeghini14} suggest that a detection of this novel critical phenomenon is within reach of current experiments on cold atoms.

The authors acknowledge financial support from the French agency ANR (Project No. 11-B504-0003 LAKRIDI) and Labex CEMPI (ANR-11-LABX-0007-01). N. C. thanks Thomas Wellens for useful discussions prior the beginning of this work.


\begin{thebibliography}{10}
\bibitem{footnote1} 
The physics we discuss in the Letter is not essentially modified for a disorder with finite correlation length.
\bibitem{footnote2} 
The condition $E=E_c$ is equivalent to the Ioffe-Regel criterion of atomic wavelength comparable to the mean free path $\ell$.
\bibitem{footnote3} 
The real part entering the definition of $P'$ is required to guarantee the unitarity of the evolution \cite{Wellens09}.
\bibitem{footnote4}
The approximation $g|\Psi|^2\ll E$ allowing us to neglect energy mixing can be quantified more precisely. For $E<E_c$ it reads $g/\xi^3\ll E$ and is thus fulfilled at any time provided $g$ is small enough. For $E>E_c$ and $E=E_c$ it respectively reads $g/(D_0 t)^{3/2}\ll E$ and $g/(D_0\ell t)\ll E$ and is thus fulfilled at long enough times.
\bibitem{footnote6} 
$\hbar/\tau_{\text{NL}}=g/\xi^{3}$ can be interpreted as the mean interaction energy in a volume $\xi^3$.
\bibitem{footnote5}
For $g=0$ this can be seen in Fig. \ref{beta}, where the slope of the theoretical black curve (giving $1/\nu$) differs from that of its numerical counterpart.
\end{thebibliography}
\end{document}